\documentclass{emulateapj}
\usepackage{apjfonts}
\usepackage{lscape}

\newcommand{\solarmass}{\mbox{${\rm M_{\odot}}$}}

\def\ltsima{$\; \buildrel < \over \sim \;$}
\def\simlt{\lower.5ex\hbox{\ltsima}}
\def\gtsima{$\; \buildrel > \over \sim \;$}
\def\simgt{\lower.5ex\hbox{\gtsima}}

\slugcomment{accepted to ApJ}

\shorttitle{Optical Spectroscopy of SS433}
\shortauthors{Kubota et al.}

\begin{document}

\title{Subaru and Gemini Observations of SS~433:\\
New Constraint on the Mass of the Compact Object\altaffilmark{1}}

\author{K. Kubota\altaffilmark{2}, Y. Ueda\altaffilmark{2},
S. Fabrika\altaffilmark{3}, A. Medvedev\altaffilmark{4},
E.A. Barsukova\altaffilmark{3}, O. Sholukhova\altaffilmark{3},\\
V.P. Goranskij\altaffilmark{5}}

\email{kaori.k@kusastro.kyoto-u.ac.jp}

\altaffiltext{1}{Based on data collected at Subaru Telescope, which is
operated by the National Astronomical Observatory of Japan. Based on
observations obtained at the Gemini Observatory, which is operated by
the Association of Universities for Research in Astronomy, Inc., under
a cooperative agreement with the NSF on behalf of the Gemini
partnership: the National Science Foundation (United States), the
Science and Technology Facilities Council (United Kingdom), the
National Research Council (Canada), CONICYT (Chile), the Australian
Research Council (Australia), Minist\'{e}rio da Ci\^{e}ncia e 
Tecnologia (Brazil) and SECYT (Argentina)}

\altaffiltext{2}{Department of Astronomy, Kyoto University,
Kyoto 606-8502, Japan}
\altaffiltext{3}{Special Astrophysical Observatory, Nizhnij Arkhyz,
Karachaevo-Cherkesiya 369167, Russia}
\altaffiltext{4}{Moscow State University, Moscow 119991, Russia}
\altaffiltext{5}{Sternberg Astronomical Institute, Moscow 119992, Russia}

\begin{abstract}

We present results of optical spectroscopic observations of the
mass donor star in SS~433 with Subaru and Gemini, with an aim to best
constrain the mass of the compact object. Subaru/FOCAS observations
were performed on 4 nights of October 6--8 and 10, 2007, covering
the orbital phase of $\phi=0.96-0.26$. We first calculate cross
correlation function (CCF) of these spectra with that of the reference
star HD~9233 in the wavelength range of 4740--4840 \AA. This region is
selected to avoid ``strong'' absorption lines accompanied with
contaminating emission components, which most probably originate from
the surroundings of the donor star, such as the wind and gas
stream. The same analysis is applied to archive data of
Gemini/GMOS taken at $\phi=0.84-0.30$ by \citet{hillwig2008}. From the
Subaru and Gemini CCF results, the amplitude of radial velocity curve
of the donor star is determined to be 58.3$\pm$3.8 km s$^{-1}$ with a
systemic velocity of 59.2$\pm$2.5 km s$^{-1}$. Together with the
radial velocity curve of the compact object, we derive the mass of the
donor star and compact object to be $M_{\rm O}$=12.4$\pm$1.9~
\solarmass\ and $M_{\rm X}$=4.3$\pm$0.6~\solarmass, respectively. We
conclude, however, that these values should be taken as {\it upper
limits}. From the analysis of the averaged absorption line profiles of
strong lines (mostly ions) and weak lines (mostly neutrals) observed with
Subaru, we find evidence for heating effects from the compact
object. Using a simple model, we find that the true radial velocity
amplitude of the donor star could be as low as 40$\pm$5 km
s$^{-1}$ in order to produce the observed absorption-line profiles. Taking
into account the heating of the donor star may lower the derived masses 
to $M_{\rm O} = 10.4^{+2.3}_{-1.9}$ \solarmass\ and $M_{\rm X} =
2.5^{+0.7}_{-0.6}$ \solarmass. Our final constraint, 1.9 \solarmass\
$\leq M_{\rm X} \leq$ 4.9 \solarmass, indicates that the compact object
in SS~433 is most likely a low mass black hole, although the
possibility of a massive neutron star cannot be firmly excluded.

\end{abstract}

\keywords{accretion, accretion disks --- stars: individual
(SS~433, V1343~Aquilae) --- supergiants --- X-rays: binaries
 --- techniques: spectroscopic }

\section{INTRODUCTION}

The microquasar SS~433 is a target of great interest in modern
astronomy as a unique Galactic source that shows steady relativistic
($v=0.26 c$) jets \citep[for a review see, 
e.g.,][]{margon1984,fabrika2004}. It gives us an ideal
opportunity to study the formation mechanism of astrophysical jets
under supercritical mass accretion onto a compact object. Although
SS~433 has been studied for about 30 years since its discovery, the
identification of the compact object, the most fundamental issue to
understand this system, has remained unsolved. In particular, the
question whether
it is a neutron star or a black hole is still open.

A direct way to identify the compact object in a binary system is
to determine its mass function by measuring the Doppler
shifts of the stars due to the orbital motion. 
In the case of SS~433, both the inclination angle and the orbital period
have been measured with high accuracy to $i=78.8\degr$
\citep{margon1989} and $P=13.082$ days \citep{goranskii1998},
respectively. Hence, if the radial velocity amplitudes of the compact
object and the donor star (or companion star) are known, their masses can
be firmly determined. 

Most of the optical light from SS~433 is emitted from the
compact object (i.e., from the jets and the accretion disk). Hence,it is
relatively easy
to measure the radial velocity of the compact object, $K_{\rm X}$,
utilizing emission lines that can be observed even with moderate-size
telescopes (hereafter, the subscripts X and O represent the compact
object and the donor star, respectively). Using \ion{He}{2} $\lambda 4686$, 
\citet{fabrika1990} obtained $K_{\rm X}=175\pm 20$~km
s$^{-1}$.
Their analysis is based on out-of-eclipse data with minimum inclination, i.e., when the
disk is oriented maximally toward us. 
\citet{gies2002b} used \ion{C}{2} $\lambda
7231, 7236$ blended
lines and determined $K_{\rm
X}=162\pm29$ km s$^{-1}$. The systemic velocity of the compact object,
$\gamma_{\rm X}$, is different between \citet{fabrika1990} and
\citet{gies2002b}, however. \citet{gies2002b} argue that this is
caused by the difference of the line emitting regions in the accretion
disk. 

The measurement of the radial velocity of the donor star, $K_{\rm O}$,
is more complicated, since the signal from the donor star,
detectable as absorption lines from its surface, is hidden in the
high flux from the compact object. \citet{gies2002a} detected faint
absorption lines in the blue part of the optical spectra of SS~433,
showing Doppler shifts expected from the orbital motion of the
donor star. They found that the spectrum resembles that of an A-type
evolved star, as confirmed by \citet{cherepashchuk2005}.  Latest
results on the radial velocity of the donor star are reported by
\citet{hillwig2008}, who used data obtained at the Kitt Peak
National Observatory (previously published in \citet{hillwig2004}) and
the Gemini telescope. 
They derived a donor star semi-amplitude of $K_{\rm O}=58.2\pm 3.1$ km 
s$^{-1}$ and a systemic velocity of $\gamma_{\rm O}=73\pm 2$ km s$^{-1}$.
Combining their results with $K_{\rm X}=168 \pm 18$ km s$^{-1}$, the average value of
\citet{fabrika1990} and \citet{gies2002b},
they conclude the mass of the donor and compact object to be $M_{\rm
O}=12.3\pm3.3$~\solarmass\ and $M_{\rm X}=4.3\pm0.8$~\solarmass,
respectively. In the discussion below (\S~6.1), we review in detail the
history of the 
radial velocity determinations of the compact object and the donor
star.

The selection of absorption lines originating from the photosphere of the
donor star is a key issue for a reliable determination of the radial
velocity of the donor star. It is also important to observe SS~433 
when the disk is oriented maximally towards the observer and the outflowing
material does not intersect with the line of sight. \citet{charles2004},
\citet{barnes2006}, and \citet{clark2007}  suggested that the
spectral type of the donor star is an A-type supergiant. However, the
wavelengths of the lines they observed did not follow the expected orbital velocity
curve. \citet{barnes2006} and \citet{clark2007} pointed out that some
absorption lines originate from the mass accretion flow onto the compact
object, and not from the surface of the donor star. More importantly,
the heating of the donor surface by the compact
object may significantly affect the accurate measurement of the radial
velocity \citep{cherepashchuk2005}.

The intensity of the absorption lines is maximum in the central
phase of the accretion-disk eclipse and it rapidly decreases when the
compact object moves out from the eclipse \citep{hillwig2004}. There
are at least two types of absorption lines in the SS~433 spectra, (1)
absorption lines with emission components, and (2) pure absorption
lines. The former ones are usually stronger, showing typical
``shell-type'' line profiles, where the absorption line is located
between two emission components. The pure absorption lines are weaker,
and they are apparently not accompanied by emission features. Hereafter, we
call these two types of absorption lines ``strong'' and ``weak'',
respectively.

The strong absorption lines have large oscillator strengths, and are
usually formed in the upper regions of a star's atmosphere. In the case of
SS~433, considering the heavy mass-loss rate of the donor of $\dot M
\sim 10^{-4}$ \solarmass\ yr$^{-1}$ \citep{fabrika2004}, and the
underlying emission components, we expect that the strong absorption
lines and associated emission lines may be formed in places not
directly related to the donor photosphere. \citet{hillwig2004}
used a ``highly rectified continuum'' (see below) to smooth out the
emission components of the strong absorption lines.
Generally speaking, 
using the strong absorption lines without
a detailed modeling of each spectral feature, has to be considered as risky.

The heating of the donor star by the strong UV radiation from the
supercritical accretion disk is known to be important. For a
disk UV luminosity of $\sim 10^{40}$ erg s$^{-1}$, the heated surface
of the donor star in SS~433 is expected to have a temperature of $\sim
20000$~K \citep{fabrika2004}, instead of only 9500~K in the
absence of heating. 
\citep{hillwig2004,cherepashchuk2005}. Such a strong heating effect
can produce the emission components observed in the strong absorption
lines. It can also distort the radial velocities measuring from the weak
absorption lines \citep{cherepashchuk2005}, since they are observed in
the non-heated (or partly heated) regions of the
donor surface, whose configuration changes with the orbital phase.

Here, we present the most recent determination of the radial velocities
and mass of the donor star and the compact object. In our analysis, we
consider various effects as discussed above. 
Firstly, we carefully
select absorption lines that originate from the donor's photosphere
with minimum contamination by emission components, such as those from
the wind, the gas stream, and the heated surface of the donor star. For this
purpose, we use high quality optical spectroscopic data obtained with
Subaru FOCAS in 2007 October during four nights. Archival data taken at
the Gemini telescope published by \citet{hillwig2008} are analyzed as
well. We present our best constraints on the mass of the compact
object in SS~433, based on a model that accounts for the averaged
absorption line profiles with consideration of the heating effects
from the compact object.

\section{OBSERVATIONS AND DATA REDUCTION}

\subsection{Subaru Data}

We observed SS~433 with the FOCAS instrument \citep{kashikawa2002} on
the Subaru telescope on October 6--8 and 10, 2007.  The jet of this
source is known to precess with a period of 162.15
days.
This epoch was chosen to observe the system in a particular phase.
Firstly, the disk was oriented maximally towards us ($\psi \approx 0$,
where $\phi$ is precessional phase), which prevented the gas outflow
from the accretion disk to intersect with the line of sight. Secondly, 
the orbital phase included the eclipse of the compact object by the
donor star ($\phi \approx 0$, where $\phi$ is orbital phase).
The spectra cover the orbital phase of $0.96 \le \phi \le 0.26$
and precession phase of $0.02 \le \psi \le 0.04$, based on the 
orbital light curves presented below and the precession ephemeris given
by \citet{gies2002b}.

FOCAS was operated with the $0'' \! .4$ slit, VPH450 Grism, and 3$\times$1
binning for the CCD chip. The sky condition was mostly photometric,
except for October 10, 2007, with a typical seeing of $\approx$
$1'' \! .0$. The resulting spectra cover the wavelength range of 3750--5250
\AA\ with a dispersion of 0.37 \AA\ pixel$^{-1}$. Per night, we
took 5 to 8 frames with 11--20 minutes exposure each.  As reference
stars we observed HD~9233 (spectral type A4 Iab), whose spectrum is
similar to the donor star in SS~433 \citep{hillwig2004}, HD~187982
(A1~Ia) and HD~332044 (B3~Ia).

All spectra are reduced by the IRAF package
\citep{tody1993} in the usual way. We first subtract the bias, using
averaged data of 20 bias frames. Then, to correct for the individual
difference of each frame, we further subtract the remaining offsets in the
over-scanned region from the exposed region.
 For the data of SS~433 with long exposures, the cosmic ray particle traces
are removed using the {\it lacos\_spec} task
\citep{vandokkum2001}. The averaged flat image is created from 13
frames taken each night, which is then corrected approximately for the
wavelength dependence of the flux to give a ``normalized'' flat
frame. Finally, we divide the object frames by the normalized flat
frame. Any remaining cosmic ray traces are removed manually in this stage.

Accurate wavelength calibration is a critical point for our scientific
goals. We utilize a Thorium-Argon lamp with the {\it identify} task on
IRAF. The accuracy is confirmed to be better than 4 km s$^{-1}$ by
checking the interstellar absorption feature of \ion{Ca}{2} $\lambda 3933.66$. 
The flux calibration is made using the standard stars, BD+28D4211
(first night) and BD+40D4032 (second-fourth night). We ignore the
effects of the slit loss, as we are mainly interested in the change of
the relative flux. Finally, the atmospheric extinction is
corrected. For the spectra of the fourth night, when the sky condition
was not photometric, we correct the flux level relative to that of the third
night by using the B-band magnitudes reported in \S~2.3. To
achieve the best signal-to-noise ratio, we add all the individual
spectra produced in this way, except for those with low statistics or
poor observing conditions, to obtain one spectrum for each night. We
utilize 5, 7, 5 and 7 frames for the 1st, 2nd, 3rd and 4th night,
respectively. We then produce ``normalized'' spectra, by
dividing the original spectra by a smooth continuum.

\subsection{Gemini Data}

We analyze the archival data of SS~433 observed with the GMOS
instrument on the Gemini telescope on June 7--13,2006 (UT). This
data was also used by \citet{hillwig2008}. The observations cover the
orbital phase of $\phi=0.84-0.30$ at the precession phase of
$\psi=0.02-0.06$, when the accretion disk is oriented maximally toward
us. In the epoch of the Gemini observations, SS~433 was found to be
more active than during the Subaru observations. Starting from the
original frames available from the Gemini web
site\footnote[6]{http://www4.cadc-ccda.hia-iha.nrc-cnrc.gc.ca/gsa/}, 
we reduced the data using the
ESO-MIDAS package \citep{warmels1992}, according to standard
procedures. Like for the Subaru data, we produce an averaged
spectrum for each night, ``normalize'' by a continuum fit, and
apply a heliocentric correction to the wavelengths.
Table~1 summarizes the observation dates and exposures of the Subaru
and Gemini data analyzed in this paper.

\subsection{Photometric data}

We obtained photometric data of SS~433 in the standard B and V bands
at the 1m telescope of the Special Astrophysical Observatory (SAO
RAS) on October 2--11, 2007 with the CCD detector EEV 42-40.  In
addition, we obtained Subaru V-band images taken just before the
spectral observation. To determined the B magnitudes of
SS~433 during the Subaru spectroscopic observations, we used the V
data and an interpolation of the (B-V) versus V relation, which is
established very well in SS~433 \citep{goranskii1998}. The final photometric
accuracy is 0.01 and 0.02 magnitudes in the V and B
bands, respectively, in direct observations. For the B-band magnitudes
interpolated to the Subaru observation time, we obtained an accuracy of
0.03 mag.

\begin{figure}
\begin{center}
\includegraphics[angle=270,scale=0.3]{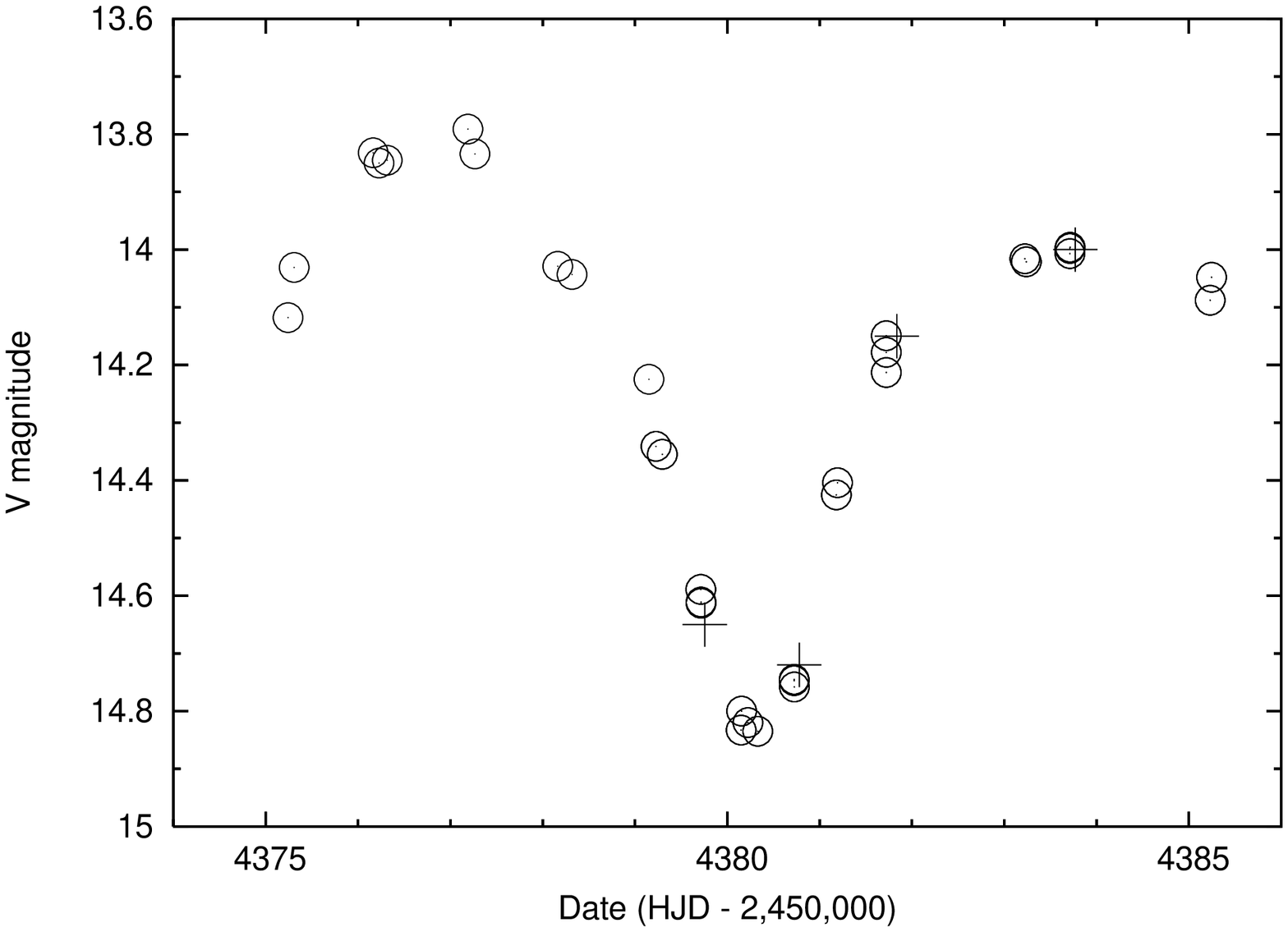}
\includegraphics[angle=270,scale=0.3]{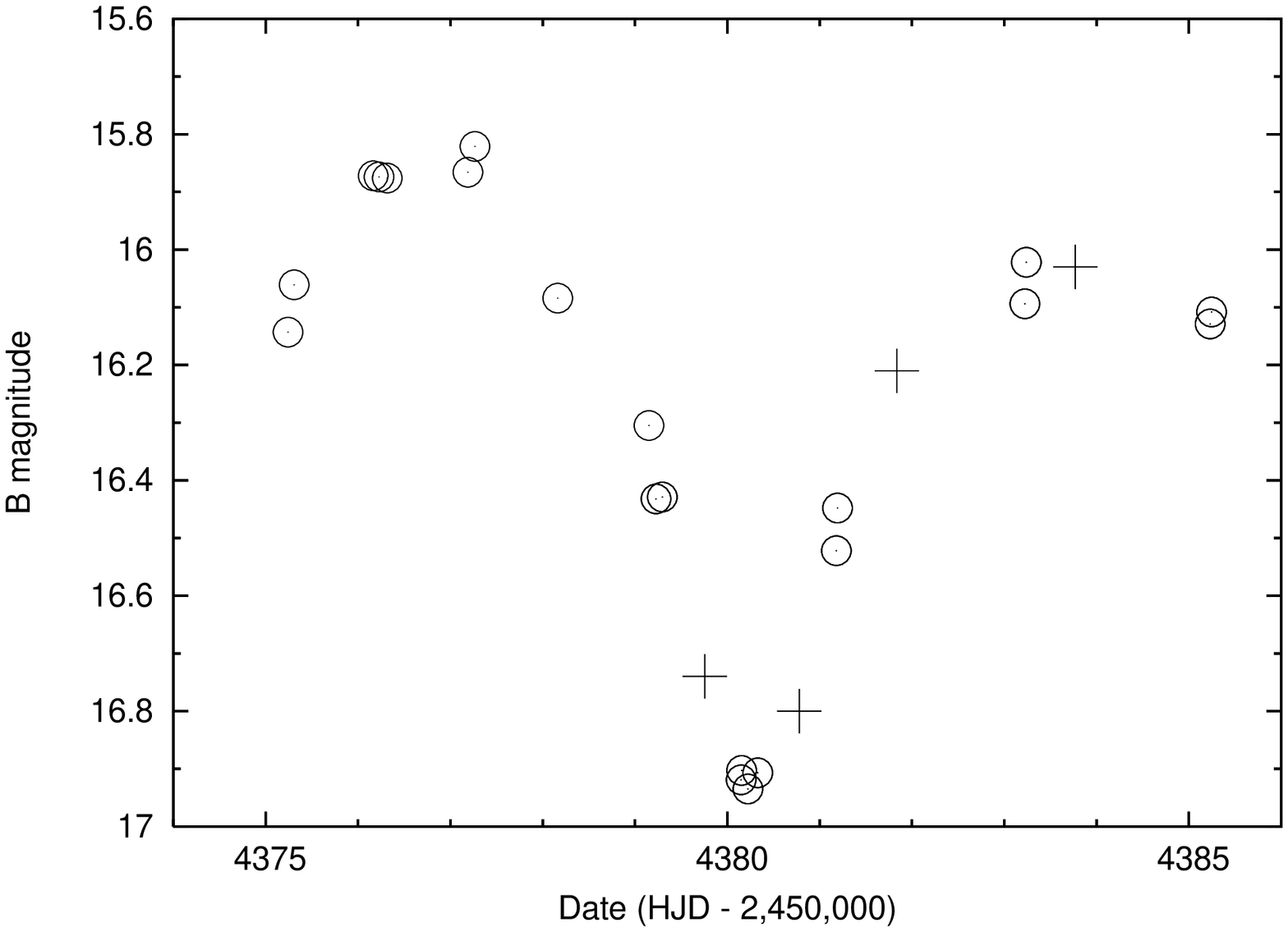}
\end{center}
\caption{Optical photometry of SS~433 during the Subaru observations.
Crosses show data points interpolated to the time of the 
spectral observations. Variations larger than 0.01 in V and 0.02 in B
represent real photometric activity in SS~433.
 \label{fig:photom}}
\end{figure}

Figure~\ref{fig:photom} shows the V and B photometric light curves of
SS~433. The SS~433 brightness out of eclipse is V = 14.0, indicating
that the object was in ``passive state'' \citep{fabrika2004}. The middle
eclipse took place between the first and the second night of the Subaru
observations. The minimum is very well shaped and regular. Using these data
and all previous photometric data of SS~433  we have,
we update the orbital ephemeris. The main minimum is 
Min\,I = JD\,$2450023.76 \pm 0.2  + (13.08227 \pm 0.00008)\times E$.
The new orbital period is slightly greater than the previous one
(13.08211) published by \citet{goranskii1998}. 
This does not mean, however, that we detect a change of the period. The
orbital ephemeris satisfy the previous photometric data as a solution
with a constant period. The particular photometric eclipse displayed in 
Figure~\ref{fig:photom} has occurred 0.375 day after the predicted
time from \citet{goranskii1998} and 0.18 day after the time predicted
by our new ephemeris. Such deviations are well-known from the
photometric behavior of SS~433.
In the following, we assume JD\,2454380.335 as the peak of the eclipse to
calculate the orbital phase. For the analysis of the Gemini spectra,
taken in June 2006, we apply the new orbital ephemeris as presented
above. 

\section{SPECTRAL FEATURES FROM THE DONOR STAR}

\begin{figure}
\begin{center}
\includegraphics[angle=270,scale=0.5]{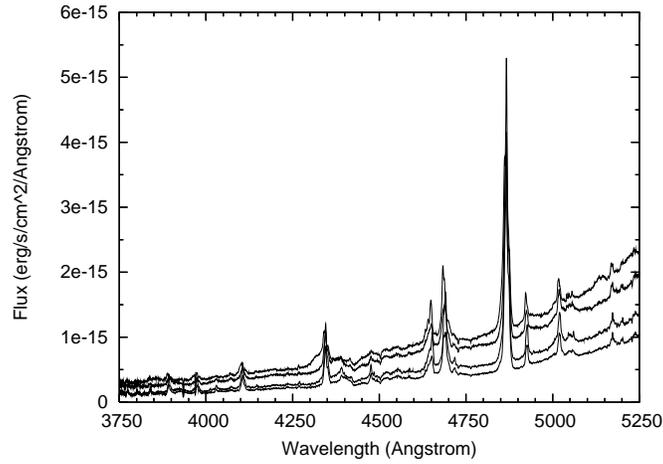}
\end{center}
\caption{Optical spectra of SS~433 observed with Subaru FOCAS
covering the 3750--5250~\AA\ range. From top to bottom, the curves
 represent the spectra taken on the fourth ($\phi=0.262$), 
third ($\phi=0.115$), first
($\phi=0.956$), and second night ($\phi=0.034$).\label{fig:all_spec}}
\end{figure}

Figure~\ref{fig:all_spec} shows the flux-calibrated spectra of SS~433
in the 3750--5250 \AA\ range obtained with the Subaru FOCAS instrument.
Apparently, the continuum fluxes were small in the first and second
night, corresponding to orbital phases close to the eclipse
($\phi=0.956$ and 0.034), and increased as the compact object
move out of the eclipse.
The most prominent features in these spectra are emission lines
originating from the accretion disk and the gas stream \citep{fabrika2004}, 
including \ion{H}{1} lines (from H$\beta$ to H$11$),
\ion{He}{1} (the strongest are 5048~\AA, 5015~\AA, 
4922~\AA, 4713~\AA, 4471~\AA), \ion{He}{2} (4686~\AA), 
\ion{Fe}{2} (the strongest is 5169 \AA), and the \ion{C}{3}\,/\,\ion{N}{3} 
Bowen blend ($\approx$ 4640 \AA). The broad lines produced by
relativistic jet were
not strong during our Subaru observations. The H$\beta^-$ line is
detected close to H$\gamma^0$ line, from $\lambda \sim 4400$ \AA\ to $\lambda
\sim 4270$ \AA\, due to the jet nutation motion.

A large fraction of the optical emission of SS~433 originates
from the compact object, 
i.e., from the accretion disk and the jet bases
\citep{fabrika2004}. When the accretion disk move out from the eclipse, 
absorption lines from the donor star become very
weak and hence careful analysis is required to study their
features. To measure the orbital motion of the donor star
(\S~\ref{sec:donor}), we need to determine the cross-correlation 
function (CCF) with
the spectrum of a reference star first. For this,
we define three different spectral regions that are not affected by
prominent emission lines from compact object; Region~1 (4490--4630 \AA),
 Region~2 (4740--4840 \AA), and Region~3 (4950--4990 \AA).

\begin{figure}
\begin{center}
\includegraphics[angle=270,scale=0.5]{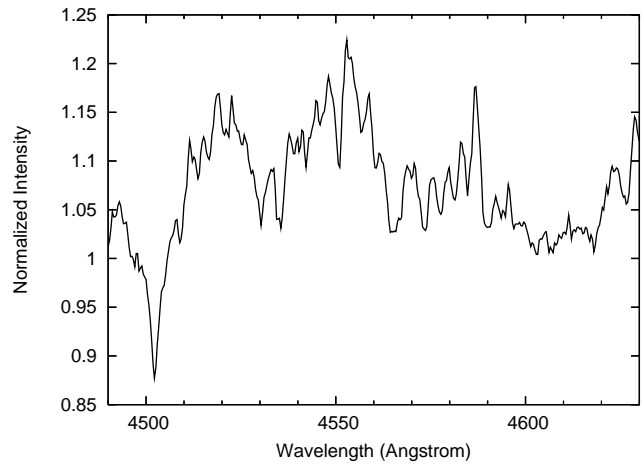}
\end{center}
\caption{``Normalized'' spectrum of Region 1 taken during the first
night of the Subaru observations. It contains several strong absorption
lines with emission components (cf. the highly rectified spectrum of
Region 1 in Figure~\ref{fig:sub_spec_region1}). The strong absorption
line at $\lambda 4501$ is a diffuse interstellar band (DIB) blended
with a \ion{Ti}{2} line.
\label{fig:region1_1sta_con}}
\end{figure}

\begin{figure}
\begin{center}
\includegraphics[angle=270,scale=0.5]{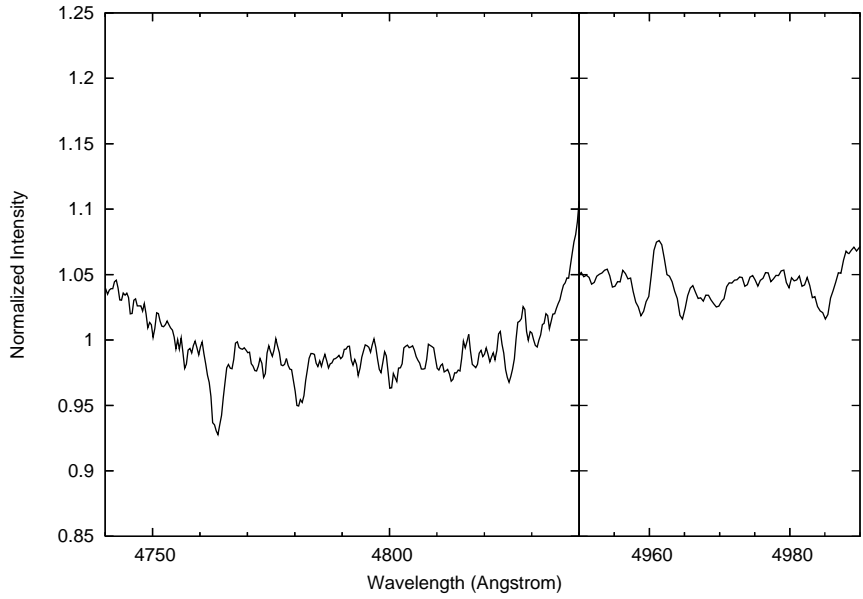}
\end{center}
\caption{``Normalized'' spectrum of Regions 2 and 3 taken on the first
night of the Subaru observations. 
\label{fig:region23_1sta_con}}
\end{figure}

The normalized spectra of Region~1 and Regions~2--3, taken during first
night, are plotted in Figures~\ref{fig:region1_1sta_con} and
\ref{fig:region23_1sta_con}, respectively. Region~1 contains many
strong absorption lines of \ion{Fe}{2} and \ion{Ti}{2} surrounded by
emission components. By contrast, Region 2 is practically void of 
strong absorption lines and contains the \ion{Cr}{2}~$\lambda 4824$ line
with a weak emission component. Region 3, the narrowest one,
contains the \ion{Fe}{1}~$\lambda 4957$ line with a weak emission
component. To make the absorption features clearly visible, we further
divide the ``normalized spectra'' by a continuum function modeled by
Legendre polynomials of order $\approx$15 in each region. We call the
resultant spectra ``highly rectified spectra''.

\begin{figure}
\begin{center}
\includegraphics[angle=270,scale=0.5]{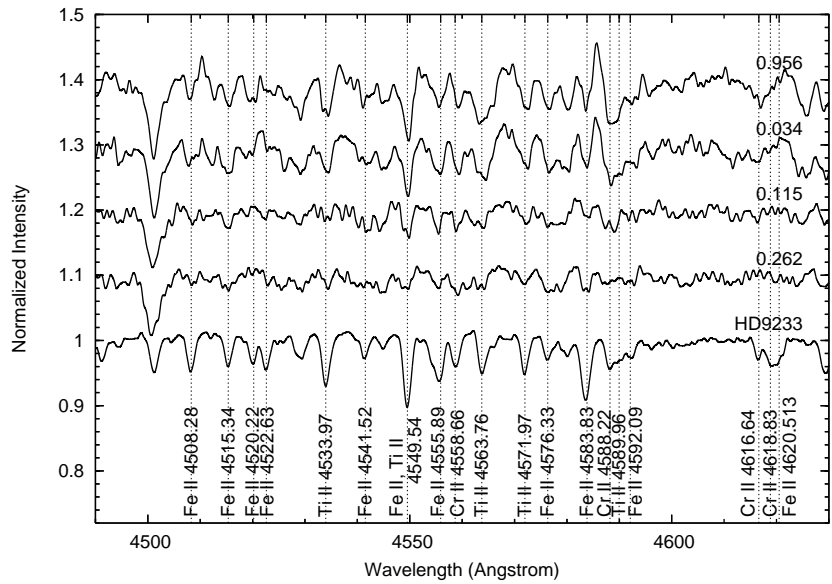}
\end{center}
\caption{Rest frame spectra of SS~433 (highly rectified) and HD~9233
(normalized) in Region~1 obtained with Subaru FOCAS.
The SS~433 spectra are shifted by offsets of 0.4, 0.3,
0.2, and 0.1 for the first, second, third, and fourth night,
respectively. Numbers
on the right side indicate the corresponding orbital phases $\phi$.
 The flux level of HD~9233 spectrum is reduced by a factor of 0.36. 
The absorption
feature near 4501.79 \AA\ is of interstellar origin.
\label{fig:sub_spec_region1}}
\end{figure}

\begin{figure}
\begin{center}
\includegraphics[angle=270,scale=0.5]{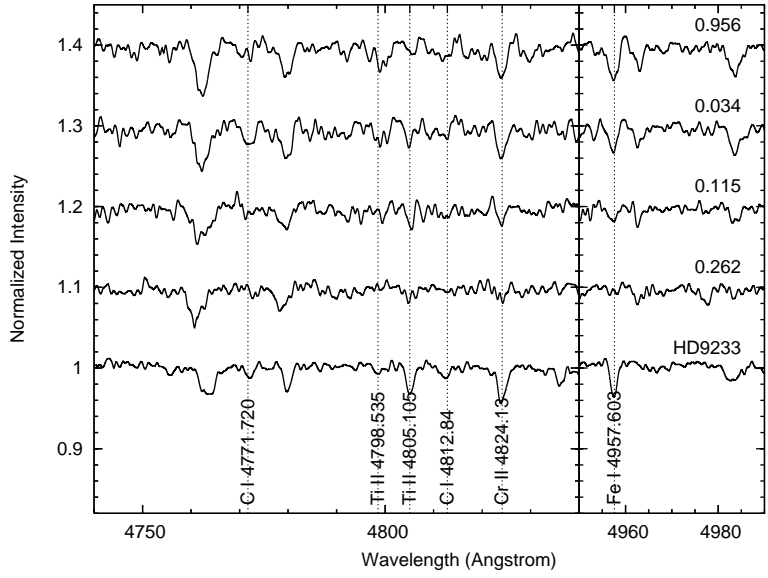}
\end{center}
\caption{Same as Figure~\ref{fig:sub_spec_region1}, but for Regions 2
and 3. The absorption features near 4762.61 \AA, 4780.02 \AA,
and 4963.88 \AA\ are of interstellar origin.
\label{fig:sub_spec_region23}}
\end{figure}

Figures~\ref{fig:sub_spec_region1} and \ref{fig:sub_spec_region23}
show the highly rectified spectra of SS~433 in Region 1 and Region
2--3, respectively, together with the normalized spectrum of the
standard star HD~9233. HD~9233, spectral type 
A4 Iab, is known to show an absorption line spectrum similar to
the donor star of SS~433 \citep{hillwig2004}. For an easy comparison,
all the spectra have been shifted into the rest frame by correcting
for its radial velocity as determined by the CCF analysis in the next
section.  It is known that even during the eclipse the surroundings of
the compact object (probably the accretion disk wind) contribute
significantly 
to the total brightness of the system. The spectrum of HD~9233 is
scaled to match the flux of the SS~433 spectra by multiplying with a
factor of 0.36 \citep{hillwig2004}. The deep absorption lines at 4500
\AA, 4760 \AA, 4780 \AA and 4980 \AA\ are due to interstellar
absorptions \citep{hobbs2008}.  Apart from these, the spectra of
HD~9233 and SS~433 contain the same set of absorption lines. These absorption
features in the SS~433 spectra become deeper as the donor star hides
the compact object \citep{hillwig2004}, providing evidence that they
originate from the donor star.

\begin{figure}
\begin{center}
\includegraphics[angle=270,scale=0.5]{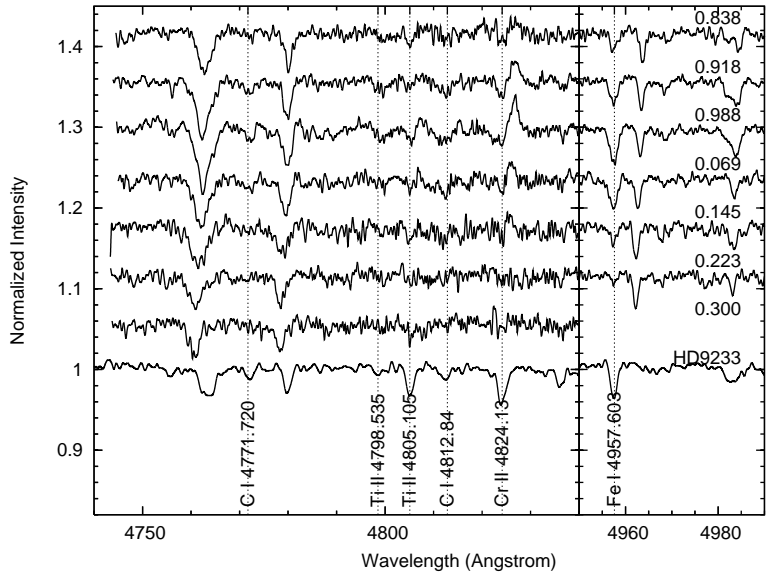}
\end{center}
\caption{Highly rectified spectra of SS~433 obtained the
Gemini GMOS in Region 2 and 3, together with the normalized spectrum
of HD~9233 obtained with Subaru FOCAS.  The absorption features
near 4762.61 \AA, 4780.02 \AA, and 4963.88 \AA\ are
of interstellar origin.
\label{fig:gem_spec_region23}}
\end{figure}

The Gemini spectra observed for 7 nights are analyzed with the same
procedure.  Figure~\ref{fig:gem_spec_region23} shows the resulting highly
rectified spectra in Regions~2 and 3 obtained with the Gemini GMOS. For
comparison, we
also plot the Subaru HD~9233 spectrum in the same figure. The
wavelengths are corrected for its radial velocity, except for that of
the seventh night where the absorption-line features are found to be
extremely faint. We do not analyze the Gemini spectra of Region~1,
since the strong absorption lines in the region appear as pure
emission lines. This is probably due to the higher activity during the
Gemini observations than during the Subaru observations. This make it
impossible to use them for our CCF analysis. 

\section{THE RADIAL VELOCITY OF THE DONOR STAR}
\label{sec:donor}

\subsection{Cross Correlation Function Analysis}
\label{sec:ccf}

We derive the radial velocity of the donor star
by cross-correlating the spectra of SS~433 with those of HD~9233 in
each spectral region (Region~1, 2 or 3). Thereby, we assume the heating effects
by the compact object at the surface of the donor star are
negligible (see discussion in \S~\ref{sec:heating}). We further ignore 
wavelengths with strong interstellar absorption. We
pay special attention to the determination of the radial velocity of
HD~9233, which is known to be a radial velocity variable star 
\citep{hillwig2008}. By measuring the Doppler shifts of
13 non-blended absorption lines \citep{chentsov2007} from the A4\,Iab
supergiant, we derive a systemic velocity of HD~9233 of
$\gamma_{\rm HD~9233}=-44.2\pm1.3$ km s$^{-1}$.  To verify our
analysis, we also measure the radial velocity of another reference
star HD~187982 (Type A1\,Ia), which was observed on October 7, 2007
(i.e., during the
second night of our Subaru observations). Our result is in good
agreement with the literature value \citep{wilson1953}
within the error range. 

\begin{figure}
\begin{center}
\includegraphics[angle=270,scale=0.5]{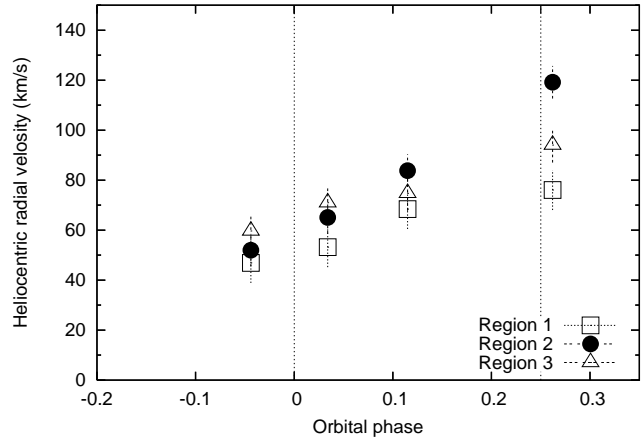}
\end{center}
\caption{Radial velocity curve of the donor star in SS~433
obtained by the CCF analysis of the Subaru data. squares
indicate the results from Region~1, filled circles from Region~2, and
triangles from Region~3.
\label{fig:sub_cross}}
\end{figure}

\begin{figure}
\begin{center}
\includegraphics[angle=270,scale=0.5]{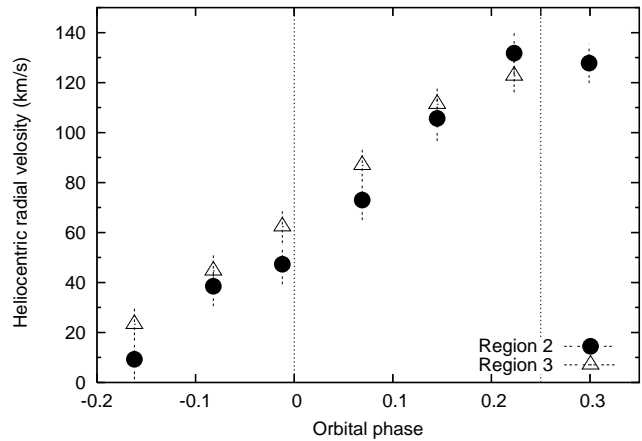}
\end{center}
\caption{Radial velocity curve of the donor star in SS~433
obtained by the CCF analysis of the Gemini data.
Filled circles and triangles indicate the results from Region~2 
and from Region~3, respectively.
\label{fig:gem_cross}}
\end{figure}

Figure~\ref{fig:sub_cross} shows the radial velocity curve of the
donor star in SS~433 from the CCF analysis of the Subaru
spectra. The error bars in each point correspond to the ``standard
error'' of the CCF \citep{fernie1989}. The amplitudes of
the radial velocity curve are different between the three spectral 
regions used in our
analysis, with Region~2 showing the largest amplitude and Region~1
showing the smallest one. In Figure~\ref{fig:gem_cross}, we display 
the CCF results for the
Gemini spectra of SS~433 together with the Subaru spectrum of HD~9233. We
confirm that the amplitude of the radial velocity curve derived from the
Gemini spectra shows the same
tendency as in the Subaru case (i.e., the amplitude derived from
Region~2 is greater than the on from  Region~3). Table~\ref{table:data}
summarized our results on the radial velocities. 

We have demonstrated that the selection of absorption lines does affect
 the estimate of the radial velocity within our simple analysis. The
differences in the amplitudes are related to the strength of the spectral
features; the absorption lines in Region 1 are the deepest and have
underlying emission components, while those in Region 2 are the
weakest and mainly do not show emission components. We interpret that
the strong absorption lines are more significantly affected by the
emission from the wind, the gas stream, and the heated surface of the donor
star, which decrease the amplitude of the radial velocity curve.
From Region~1, we obtain $K_{\rm O}=24\pm9$ km s$^{-1}$ with
a systemic velocity of $\gamma_{\rm O}=52\pm6$ km s$^{-1}$. This value
 for $K_{\rm O}$ is even smaller than the result from
 \citet{hillwig2004}, $K_{\rm O}=45\pm6$ km s$^{-1}$, 
who studied the same spectral
region. This is probably due to different conditions of the surroundings
 between the two epochs of observations.

\begin{figure}
\begin{center}
\includegraphics[angle=270,scale=0.5]{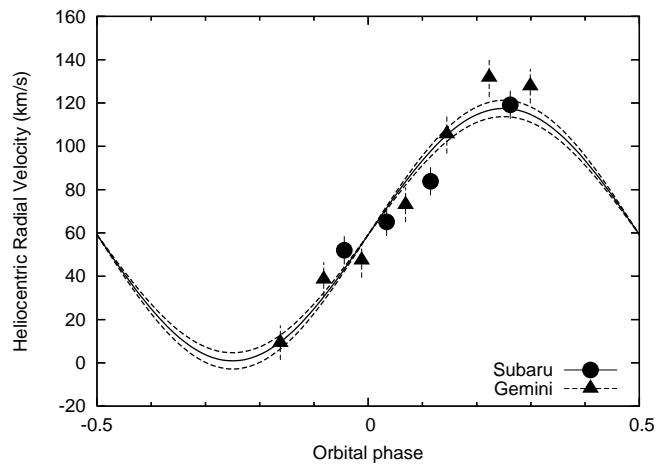}
\end{center}
\caption{Radial velocity curve of the donor star in SS~433
complied from both Subaru and Gemini results of the CCF analysis in
Region~2. Filled circles and triangles represent Subaru and Gemini
results, respectively. The best-fit Keplerian solution is displayed by
the solid curve. Dashed lines correspond to the $1\sigma$ error.
\label{fig:fit}}
\end{figure}

In this context, the selection of ``weak'' lines is important to
determine correctly the motion of the donor star,
{\it as for as the region responsible
for the production of the absorption lines is constant over the orbital
phase}. Under this assumption, we can estimate the amplitude of the
radial velocity curve by fitting the velocities with a Keplerian
solution. Figure~\ref{fig:fit} shows the Subaru and Gemini results
obtained from the CCF analysis of Region~2 together with the best-fit
curve. We
obtain a semi-amplitude of the radial velocity of 
$K_{\rm O}=58.3\pm3.8$ km s$^{-1}$ and a systemic velocity of 
$\gamma_{\rm O}=59.2\pm2.5$ km s$^{-1}$. 
This value of $K_{\rm O}$ is consistent with
the result of \citet{hillwig2008} within the error bars.
We note that Figure~\ref{fig:fit} may indicate a distortion of the
donor's radial velocity curve in the orbital phases 0.0--0.15.

\subsection{Average Absorption Lines Profiles}

The high quality of the Subaru spectra allows us to study individual
absorption lines. In the case of CCF analysis (\S~\ref{sec:ccf}), one
compares two stars (i.e., of SS~433 and reference stars) with almost
identical spectra. Hence, the complex blending and crowding of the
absorption lines is not critical for the study. The analysis of the
individual lines, however, depends strongly on such line blending
effects and requires knowledge of the ``laboratory'' wavelengths of the
blends.
We carefully check the whole
spectrum of SS~433 for the first and second night, when the system was
in maximum eclipse. First, we select two groups of absorption lines:
``strong'' lines, which clearly show emission components, and ``weak''
lines with pure absorption line profiles. For the line identification,
laboratory wavelengths and relative line strengths, we refer to the
Atomic Spectra 
Database\footnote[7]{National Institute of Standards and Technology;
http://physics.nist.gov/PhysRefData/ASD/} and to the Atomic Line
List\footnote[8]{Department of Physics and Astronomy, University of
Kentucky; http://www.pa.uky.edu/$^\sim$peter/atomic/}.  We
estimate the effective wavelengths of the blends by weighing the
wavelengths of the individual lines according to their line strengths. 
We test each line or line blend to have the same radial velocity in
the given Subaru night, allowing a difference of up to 20 km
s$^{-1}$. The set of strong lines as well as the set of weak lines are
both formed from
lines that have same radial velocities within the first and 
the second night. The data of the other nights are not
considered at the line selection. Obvious or resolved line blends are
not included in the two line groups.

Finally, we add the line profiles for each group in the normalized
spectra in the
radial-velocity space to create the strong and weak average line
profiles. In the averaging procedure, we apply a weight of unity for
a single line and a smaller weight for obvious (but non-resolved)
blends. The line blending may distort the line profiles. The averaging
procedure minimizes this distortion, because the line blending is only
occasional, and because obvious and strong blends are excluded from the
set of lines.

To derive the average line profiles, we do not use the ``highly
rectified'' spectra. Instead, we apply a linear continuum rectification to
the final average line profiles in order
to subtract the continuum levels near the
lines. 
Both the ``strong'' and the ``weak'' absorption lines are considerably
fainter than the strong emission lines in the SS~433
spectrum. Hence, in case where an absorption line is located near the wing of a
strong and broad emission line, its local continuum becomes not
flat. Applying a linear function for the continuum rectification of
 the average spectra makes it possible to obtain a flat normalized 
continuum for the average line profiles. Using our B-band photometry results
(Figure~\ref{fig:photom}), we scale the average line profiles to a
relative intensity unit, where the contribution from the donor star
can be directly compared for the four nights. In the scaling process,  
we multiply the average profiles by coefficients depending on the 
system brightness and adopt the first Subaru night coefficient as 1.0.

We include 8 individual lines in the average strong line profile 
(\ion{Mg}{2} $\lambda 4481.21$,
\ion{Ti}{2}+\ion{Fe}{2} $\lambda 4549.63$, 
\ion{Ti}{1}+\ion{Fe}{2} $\lambda 4555.49$,
\ion{Fe}{2} $\lambda 4576.39$,
\ion{Ti}{2}+\ion{Fe}{2} $\lambda 4583.41$,
\ion{P}{2}+\ion{Cr}{2} $\lambda 4823.84$,
\ion{Si}{2} $\lambda 5041.03$,
\ion{Fe}{1} $\lambda 5226.86$) 
and 8 individual lines in the average weak line profile
(\ion{Cr}{1} $\lambda 4161.42$,
\ion{Fe}{1} $\lambda 4271.76$,
\ion{Ti}{2} $\lambda 4290.22$,
\ion{Ti}{1} $\lambda 4325.13$,
\ion{Fe}{1} $\lambda 4528.87$, 
$\lambda 4983.85$, 
$\lambda 5125.11$,
\ion{Mg}{1} $\lambda 5183.60$). 
We find that nearly all strong absorption lines with emission
components are generated by ions, while most of the weak, pure
absorption lines are produced by neutral atoms. This implies that the ion
absorption lines are partial formed in a more extended gas envelope of the
donor star (i.e., the donor's wind). This wind could be  a low velocity
wind, since the donor
overfills its Roche lobe. The emission components in the strong
absorption lines may be partly formed in the gas stream (probably 
the mass
flow,  see below), which is best observed in 
hydrogen and \ion{He}{1} emission lines 
\citep{crampton1981,fabrika1997,fabrika2004}.

The average strong and weak line profiles are shown in the right hand
panels of 
Figures~\ref{fig:obs_mod_s_lines} and \ref{fig:obs_mod_w_lines},
respectively. The strong lines show clear
evolution of their emission components over the four nights of
observations. The absorption components
shift with time to positive velocities with an amplitude of
$\approx 40$ km s$^{-1}$. The weak lines show the same
systematic evolution with orbital phase, although its orbital
shift in radial velocity is notably larger than that of the strong
lines. The final signal-to-noise ratio in the average line profiles is
very high. We detect several features in the line profiles that change
from night to night in ways that are difficult to interpret. They would
be due to 
line-emitting and -absorbing regions in the system having a complex
structure. Additionally, emission components could partially fill the
absorption profiles of the weak lines as well. In the following, 
we study the main features only, such as the line positions and intensities.

We measure the absorption line positions for the average weak and
strong line profiles and compare them with the CCF results. Note that
the CCF analysis and the study of the average absorption line profiles are
independent. We find that the weak absorption lines show the same
behavior as the lines in the CCF analysis of Region 2, which is free
from strong lines. The total radial velocity amplitude measured
between the first and 4th spectrum is $\approx 63$ km s$^{-1}$, which
is very close to $\approx 67$ km s$^{-1}$ derived from the CCF
analysis (Table~\ref{table:data}). The absolute velocities 
are about the same as well, with the
difference between the two methods being within 5 km s$^{-1}$. We further
compare the behavior of the strong absorption lines with the CCF
results of Region 1, which mainly contains strong lines with emission
components. The difference is 10 km s$^{-1}$ for the total radial
velocity amplitude, although the systemic velocity is larger by 15 km
s$^{-1}$ for the average strong line profile than for the CCF analysis. 
Note that for the CCF analysis, we used the 
``highly rectified'' spectra, where smoothing of the emission
components may produce a systematic shift in the absorption line
position. In the following, 
we study the average line profiles using a simple
model of a close binary by taking into account the heating effects
from the compact object.

\subsection{Study with a Simple Model Including Heating Effects by the Compact Object}
\label{sec:heating}

\subsubsection{Model Description}

\begin{figure}
\begin{center}
\includegraphics[angle=270,scale=0.6]{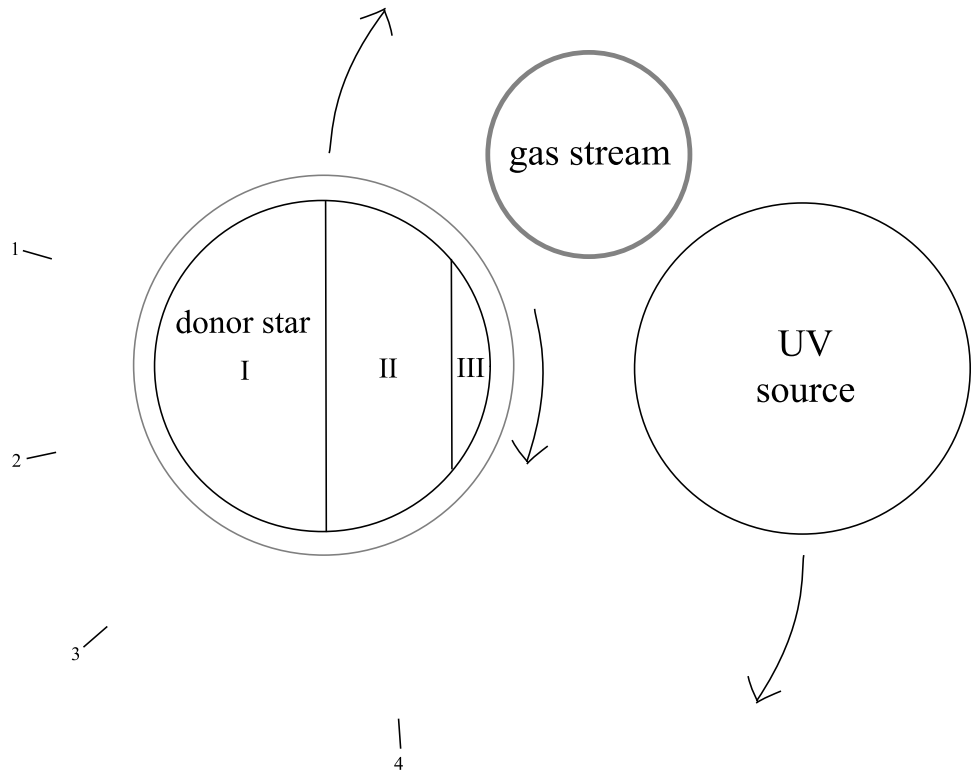}
\end{center}
\caption{
Sketch of the binary system SS~433 (not to scale). The system consists of 
a mass donor
star with an extended gas envelope, a UV heating source (surrounding
the compact object), and a gas stream. The orbital motion
and rotation of the donor are shown. We distinguish between three
 different regions of the
donor surface, (I) the non-heated region from which we observe 
absorption lines only, (II) the heated region from which we observe
emission lines only, and (III) the overheated region which does not emit
any spectral lines. The extended envelope (i.e., the wind) produces 
emission lines only in Region II and in those parts of Region I that
are exposed to the UV source. The orbital phases seen by the observer 
in the four Subaru nights are indicated.
}
\label{fig:scheme}
\end{figure}

Figure~\ref{fig:scheme} sketches the binary system with its main
components. Since the UV luminosity of the compact object is as high as 
$L_{UV} \sim 10^{40}$ erg s$^{-1}$\citep{fabrika2004}, the compact
object can heat the
donor surface up to $\sim 20000$ K during the 13-day binary period. 
Although the detailed of geometry of the system is unknown, we know that the
size of the optical continuum source is about the size of the donor-star
or slightly exceed it, because optical eclipses are never
total (about half of the continuum light remains present always).
The donor star has an extended and dense envelope, which is
probably due to a strong, low-velocity wind. Studies of X-ray eclipses in
SS~433 \citep{filippova2006} revealed that the radius of the envelope
that is opaque to X-rays exceeds the donor radius (or its Roche lobe
radius) by 10--20\,\%. This proves the existence of a gas envelope
(``coat'') around the donor star, which can produce the emission
components observed in the strong absorption line profiles. This gas
envelope is also sketched in the figure.

We construct a simple toy model of the system to study the
heating effect on the absorption lines. Heating
effects increase the observed radial velocity amplitude of absorption
lines \citep{antokhina2005,cherepashchuk2005}, 
because, due to the spin of the donor star, the non-heated side
moves with a larger velocity than that of the center of mass. In out
model, we consider three different regions of the donor surface
(Figure~\ref{fig:scheme}): Region~I, which is not heated and produces
absorption lines only, Region~II, which is heated and produces
emission lines only, and Region~III, which is overheated and therefore 
does not produce any
emission or absorption lines of elements with low ionization potentials
(like, i.e., \ion{Ti}{2} or \ion{Fe}{2}). 

We assume that the system is synchronized and the donor star is a
sphere with a volume equal to that of the Roche lobe for a given mass
ratio $q = M_{\rm X} / M_{\rm O}$. The orbital inclination of the system
is adopted as 
79$^\circ$ \citep{fabrika2004}. 
The semi-amplitude of the compact object is set to $K_{\rm X}=160$ km
s$^{-1}$, as derived late in this paper (\$~\ref{sec:compact}). The
semi-amplitude of the donor star $K_{\rm O}$ (or the mass ratio $q$) is
taken as free parameter in this model. The donor
surface is divided into 100 grid cells both in longitude and
latitude, with each grid cell producing a Gaussian absorption line profile
in the non-heated region or an emission line profile in the heated region.
A Gaussian line width of FWHM~$=5$\, km s$^{-1}$ is adopted. For
absorption lines formed at the donor's surface which is not exposed to the UV
source (Region~I in Figure~\ref{fig:scheme}), the Gaussian line
intensity is normalized to unity. For emission lines formed in the
heated region of the surface (Region~II in Figure~\ref{fig:scheme}), we
calculate the normalization depending on the heating parameters (see
below). We adopt a quadratic limb-darkening law \citep{kallrath1999}
with $x = y = 1$ for absorption lines only. Using this limb-darkening
law, 
we can easily fit the observed absorption lines. Note, however, that
determining the correct limb-darkening law for this supergiant with
heavy mass loss and heating is a truly complex task. In the case of
emission lines, we do not account
for a limb-darkening effects, since one expects
an inverse temperature gradient in the emission region.  

The gas envelope is modeled in a similar way than the donor surface,
with the difference that it produces emission lines in the heated region 
(Region~II), but
no emission or absorption lines in the other two regions. The radial
extent is set to 10\,\% of the donor radius. It has 10 individual
segments in the radial direction. We assume that the
emitting gas in the envelope rotates with the same velocity as the
donor surface and moves in radial direction with the escape velocity
$V_{\rm esc}$. 

The UV source is spherical and has the same size as the donor
star in the model (Figure~\ref{fig:scheme}).
 Each point of the donor
surface sees all the points of the extended source visible from it.
We calculate the angle of incidence of the UV radiation in each point
of the donor surface. Thus, we specify only geometrical properties of
the source. We suggest that the donor's regions exposed to the UV
radiation of the source produce emission lines. 
This is expected if the temperature
gradient in the donor's atmosphere is inverse. \citet{antokhina2005}
confirmed  this behavior in their X-ray heating model of a low-mass X-ray
binary. UV radiation is subject to strong extinction and may therefore
not penetrate deeply into the donor atmosphere.
Considering the high luminosity ($L_{UV} \sim 10^{40}$\,erg/s)
of the accretion disk in SS~433, however, we suggest that the UV
radiation can indeed produce the inverse temperature gradient in the
atmosphere and that the gas in the
donor's wind may be ionized down to the photosphere.

We introduce two empirical parameters, the 
heating coefficients $\gamma_{\rm phot}$ and $\gamma_{\rm env}$,
which define the intensities of the emission lines formed in the heated
regions of the donor surface and in the envelope, respectively. 
From the relative fluxes of the incident radiation in each point
exposed to the UV source, we calculate the expected emission-line
components. We then determine the values of $\gamma_{\rm phot}$ and
$\gamma_{\rm env}$ by comparing the line profiles between 
the model and the data observed in all four nights. The values are 
independent
for the strong and the weak lines (in particular, $\gamma_{\rm env} =0$
for the weak lines). 
The obtained heating coefficients are relative ones only, and cannot be used 
to estimate the heating effects physically. To produce a reasonable agreement 
between the observed and the modeled line profiles, $\gamma_{\rm phot}$
has to be 2--3 times larger for the weak lines than for the strong lines. 
The weak lines do not require an additional emission component formed
in the envelope. In case of the strong lines, this 
component ($\gamma_{\rm env}$)
is necessary, since it produces emission line wings which are notably
broader than the photospheric line profiles.
By varying other parameters of the model, such as the mass ratio,
we can infer the required amount of heating that is necessary to account
for the observed line profiles.

Finally, we integrate the line profiles from individual regions of the 
donor surface and of the envelope which are visible from the observer
during the
orbital phases of the Subaru observations (indicated in
Figure~\ref{fig:scheme}). The final line profiles are convolved with
the instrumental response, which is derived from single line measurements
in the comparison-lamp spectra. The normalization of the final absorption 
lines is determined from the data of the first night only.
These
normalization coefficients are kept constant for all four nights of
the observations.
 
The final line profiles are notably broader (FWHM~$\sim 100$\,km s$^{-1}$) 
than our Gaussian lines formed in the individual surface cells and they 
do not depend on the adopted line width of the individual lines
(when it is less than $\sim 20$\,km/s). 
This simple approach is sufficient for the following study, because we 
do not compare in
detail the observed and calculated line profiles, but investigate the
main features of the heating effects only.

Several effects are not taken into account here. In reality,
heating by the UV radiation is a complex process. For example, the UV
photons may not reach the donor surface because of strong
absorption in the wind. The UV absorption can heat the gas deeply down
to the surface, however.
We further assume isotropy
of the UV source, which is probably are oversimplification. For instance, the
thick outer rim of the disk \citep{filippova2006} may cast an extended
shadow. 
This effect is most important in the fourth Subaru night, since,
at precession phases $\psi \sim 0$, the donor star crosses
the disk plane at $\phi \approx$0.25 and 0.75. 
The assumption of isotropic UV radiation will lead to an overestimate of
the heating effects in this case. Since the environment of the compact object
(i.e., the wind, the jet bases and the disk structure) are basically unknown,
a complex modeling of the heating is
problematic. Thus, in this paper, we 
illustrate how the heating distorts the
radial velocities of the absorption lines in SS~433 for a better
understanding of the principal difference between the strong and weak lines.

The gas stream is a strong source of hydrogen and \ion{He}{1} emission
lines. \citet{crampton1981} showed that the radial velocity curves
have the largest redshifts at orbital phases $\sim 0$ close to the
inferior conjunction of the donor star; their orbital phases lag
behind the accretion-disk phase by 0.2--0.25. Later studies of
the hydrogen and \ion{He}{1} emission lines
\citep{kopylov1989,fabrika1997,goranskii1997} revealed that H$\beta$
and \ion{He}{1} radial velocities show the largest redshifts at
orbital phases 0.9--0.95 and 0.85--0.9, respectively. These allows also
detected a partial eclipse in the hydrogen emission lines at orbital
phases 0.1--0.2. This indicates that the \ion{He}{1} and hydrogen
emission lines in the SS~433 spectra are formed in the gas stream
onto the accretion disk.

\citet{goranskii1997} and \citet{gies2002b} also discussed whether the
behavior of the hydrogen and \ion{He}{1} radial velocity curves may 
result from an evacuation of the disk wind surrounding the donor star,
which leads to anisotropic wind and the observed radial velocity curves.
This cannot explain, however,
both the partial eclipses and the differences between the hydrogen and
\ion{He}{1} radial velocity curves. If hydrogen and \ion{He}{1} lines
are formed in the anisotropic wind, their radial velocity amplitudes
have to be greater than that of the accretion disk 
(traced by \ion{He}{2} line), since
the accretion disk powers the wind. This does not agree with the
observed radial velocity amplitudes.
We therefore conclude that the hydrogen
and \ion{He}{1} emission lines  are formed in the gas stream, although a
fraction of this emission may be formed in the disk wind as well.
In any case, the emission region must be extended. A probable
location of the gas stream region is shown in Figure~\ref{fig:scheme}.
It is noteworthy that a fraction of the emission of the strong
absorption lines
(Figures~\ref{fig:obs_mod_s_lines}) may be formed at the same location
as the hydrogen and \ion{He}{1} lines.

During all for nights, we detected an additional red emission line component in
the strong and even in the weak absorption lines,
which we fail to reproduced with our model. 
We ascribe this component
to the gas stream, which contributes stronger to the red emission
in the strong lines than in the week lines. We
do not model any probable eclipses of the gas stream
\citep{kopylov1989,fabrika1997,goranskii1997}, which may change the
intensity of the red emission components. To model this additional red
emission, we decide to follow the radial velocity curves of hydrogen
and He\,I emission lines by \citet{fabrika1997}, where the lines show
the largest redshifts (100--150~km s$^{-1}$) at the orbital phases
$\sim$0.85--0.95 and the velocity decreases gradually with the orbital
phase. Finally, the gas stream is modeled to produce Gaussian emission
lines, whose parameters are tuned to reproduce the observed line
profiles.  

\subsubsection{Comparison with the Data}

\begin{figure}
\begin{center}
\includegraphics[angle=0,scale=0.7]{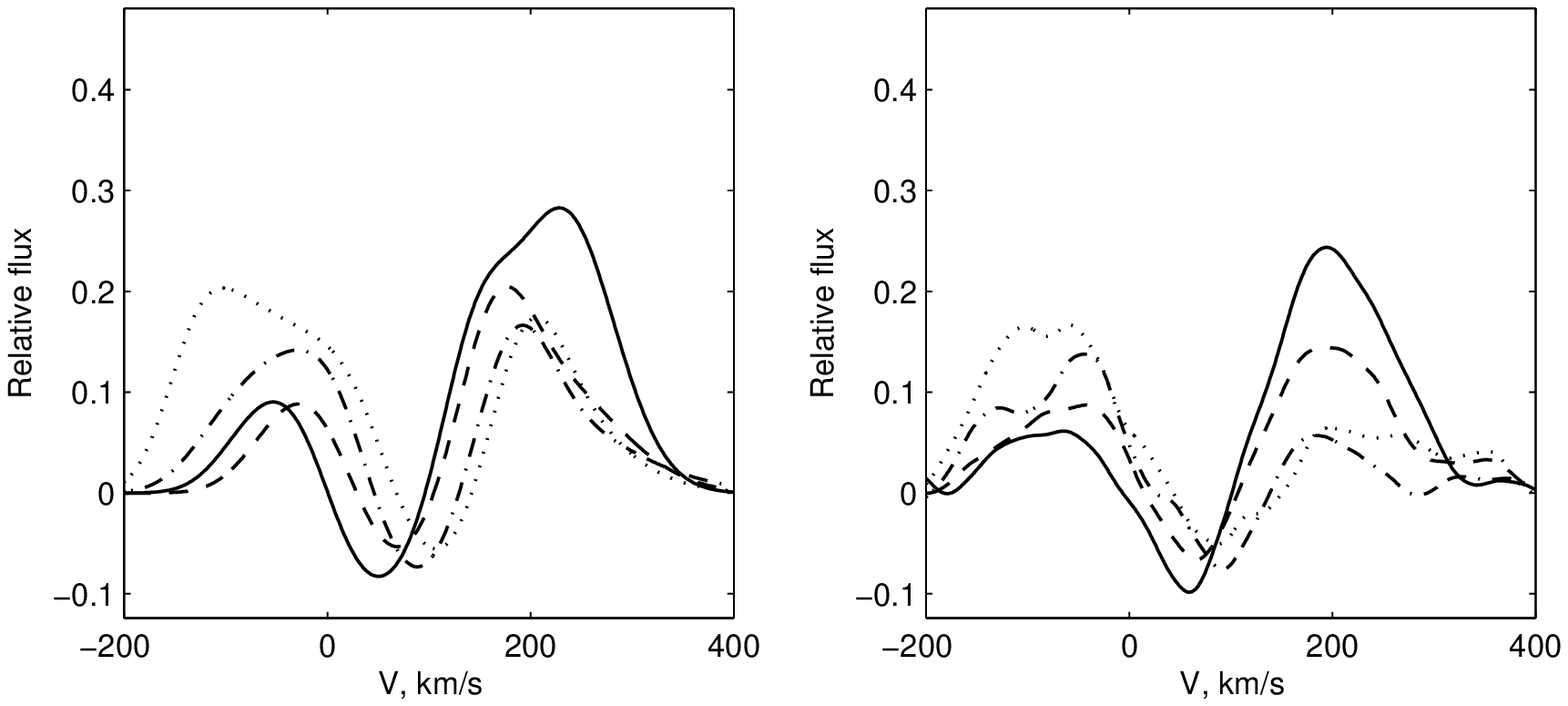}
\end{center}
\caption{
Observed average absorption-line profiles for the ``strong'' lines
with emission components (right) and its best-fit model
(left). From the first to the fourth Subaru night, the profiles are
denoted by solid, dashed, dash-dotted and dotted lines.
}
\label{fig:obs_mod_s_lines}
\end{figure}

\begin{figure}
\begin{center}
\includegraphics[angle=0,scale=0.7]{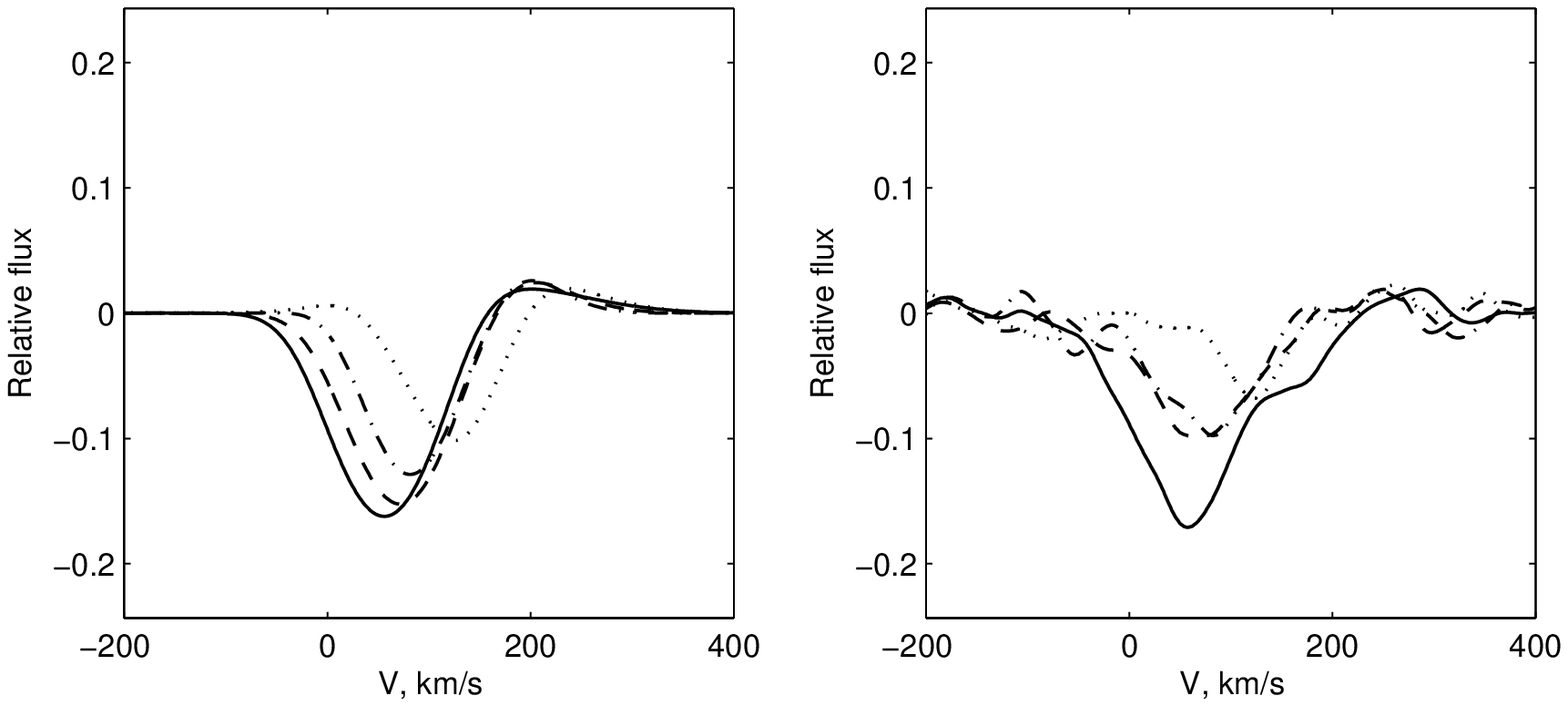}
\end{center}
\caption{
Observed average absorption-line profiles for the ``weak'' lines
without notable emission components (right) and its best-fit model
(left). From the first to the fourth Subaru night, the profiles are
denoted by solid, dashed, dash-dotted and dotted lines.
}
\label{fig:obs_mod_w_lines}
\end{figure}

The left hand panels of Figures~\ref{fig:obs_mod_s_lines} and
\ref{fig:obs_mod_w_lines} show the best-fit models of the strong and
weak line profiles, respectively, to be compared with the observed
ones in the right panels. 
These line profiles are modeled using a radial velocity
amplitude of the compact object of $K_{\rm X}=160$ km s$^{-1}$ and the
real radial velocity amplitude of the
donor star $K_{\rm O}=40$ km s$^{-1}$ (i.e., $q=0.25$). The relative
heating coefficient $\gamma_{\rm phot}$ for the weak lines is twice
as large than that of the strong lines, and the wind velocity is set to $V_{\rm
esc}=260$ km s$^{-1}$ (for the strong lines only).
We see that our model can reproduce the overall features of the
observations. 
For the strong absorption lines, the emission components
evolve in agreement with the idea that the gas envelope, which rotates
with the same velocity as the donor star, is heated by the
compact object.

For the weak absorption lines, emission lines originating from the heated donor
surface are required, while those from the envelope are not. 
These emission lines are seen as a low intensity emission bump near the
absorption line (as for the strong lines). This emission
component alters the position and intensity of the absorption lines.

The absorption lines move in accordance with the orbital phase, and
their velocity amplitudes are smaller for the strong lines than for the
weak lines. The absorption line intensities generally decrease with
orbital phase because of the heating (note that the observed
spectra are scaled from the photometric data in order to keep the 
non-illuminated continuum 
radiation of the donor constant). Naturally, the observed 
absorption line profiles are more complex than those produced by
our simple model. For example, the weak absorption line profile of the
first night (solid line in Figure~\ref{fig:obs_mod_w_lines}) shows
either an additional absorption in its red wing or an additional emission
component. 
Such a feature is not produced in our model, since 
we do not take into account a possible absorption of the
continuum radiation from the compact object in the donor's wind. 
This effect might be
important during the first and second night (see
Figure~\ref{fig:scheme}). 
Additional absorption features may be present for the second and
third night in the blue wing of the strong and weak absorption
lines (Figures~\ref{fig:obs_mod_s_lines},
\ref{fig:obs_mod_w_lines}). During these orbital phases, the speculated
wind from the donor, which is seen projected onto the strong continuum
source, is directed toward
us because of the stellar rotation. This feature may also produce the
distortion of radial velocity curve observed in Figure~\ref{fig:fit}.

We conclude that the model reproduces both the intensity and the radial 
velocity
variations of the emission components and the absorption lines. 
The modeled line
profiles are influenced mostly by the following parameters; the real 
velocity amplitude of the donor star
$K_{\rm O}$, and the heating efficiency coefficients $\gamma_{\rm phot}$
and $\gamma_{\rm env}$. As mentioned above, the best-fit value is 
$K_{\rm O} \approx 40$ km s$^{-1}$ (i.e., $q \approx 0.25$ with $K_{\rm X}
\approx 160$ km s$^{-1}$). This is 18 km s$^{-1}$ or 30\,\% less than
that measured in our CCF analysis (58.3$\pm$3.8 km s$^{-1}$).
It is impossible to produce the average line profiles in our model
for $K_{\rm O} < 35$ km s$^{-1}$. 
Further, for velocity amplitude of the compact object between 170 and 150 km
s$^{-1}$, we obtain values of $K_{\rm O}$ between 35 and 45
km s$^{-1}$.
We thus conclude that the donor's real radial velocity amplitude is
$K_{\rm O} = 40 \pm 5$ km s$^{-1}$, based on our simple model.

From the set of models reproducing the observed averaged line profiles,
we find that the size of the overheated region (Region III in
Figure~\ref{fig:scheme}), modeled as a cone with origin at angle of the
donor, has a half-opening angle of  $\approx$15--20$^\circ$ for the
strong lines and
$\approx$20--25$^\circ$ for the weak lines, respectively.
The wind velocity of the donor is $V_{\rm esc} \sim 260$ km s$^{-1}$. 
The mass flow, which produces parts of the red emission components
in the absorption lines, has a radial velocity decreasing from 160 km s$^{-1}$ 
in the first night to 90 km s$^{-1}$ in the last night for the strong 
absorption lines, and from 100 km s$^{-1}$ to 80 km s$^{-1}$ for the
weak also lines, respectively.
The emission line components formed in the gas stream
are broad with a FWHM of 140--200 km s$^{-1}$. The intensity of this 
component is two times weaker in the weak absorption lines than 
in the strong absorption lines. The origin and formation of the gas stream
(mainly observed in the hydrogen and \ion{He}{1} emission lines) were 
discussed in previous papers \citep{crampton1981,fabrika1997,fabrika2004}.
It is important to note that the introduction of the red emission components
in the modeled line profiles is necessary, although its formation
remains unclear. 
All these parameters do not change, however, 
the modeled absorption line profiles so strongly as the real 
radial velocity amplitude of the donor and heating efficiency coefficients do.

\begin{figure}
\begin{center}
\includegraphics[angle=0,scale=0.5]{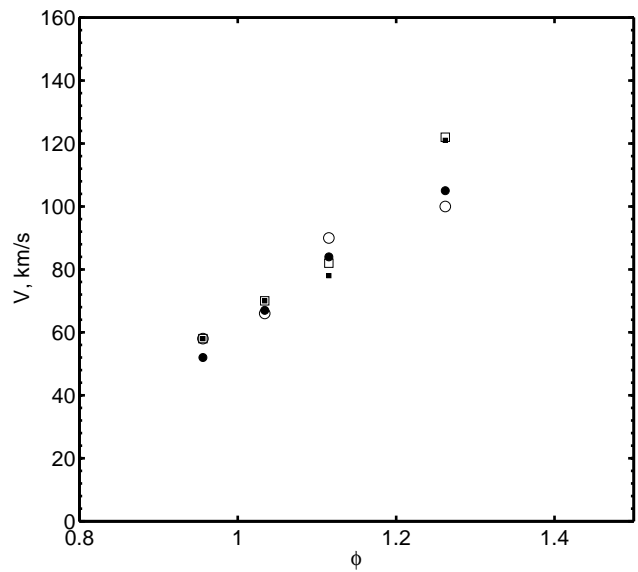}
\end{center}
\caption{
Absorption-line radial velocity curves derived from the observed
average line profiles (filled symbols) and those measured from the
best-fit models using the same method as for the observed ones (open
symbols). Circles represent the strong lines, and squares the weak
 lines, respectively. $K_{\rm O} = 40$ km s$^{-1}$ and $K_{\rm X} = 160$
km s$^{-1}$ are assumed. The heating model well reproduces the
observed radial velocities.
}
\label{fig:model_velocity_curve}
\end{figure}

Finally, in Figure~\ref{fig:model_velocity_curve}, we present the radial
velocity curves derived from the observed average line profiles,
simply based on the position of the absorption line minimum. 
For comparison, the values from our best-fit models 
($K_{\rm O} = 40$ km s$^{-1}$
and $K_{\rm X} = 160$ km s$^{-1}$) are displayed as well. 
Clearly, the heating model
reproduces well the observed radial velocities. Note that these velocities
are apparent ones and differ from the real radial velocities of the donor
star considered in the model.

\section{RADIAL VELOCITY OF THE COMPACT OBJECT}
\label{sec:compact}

\begin{figure}
\begin{center}
\includegraphics[angle=270,scale=0.5]{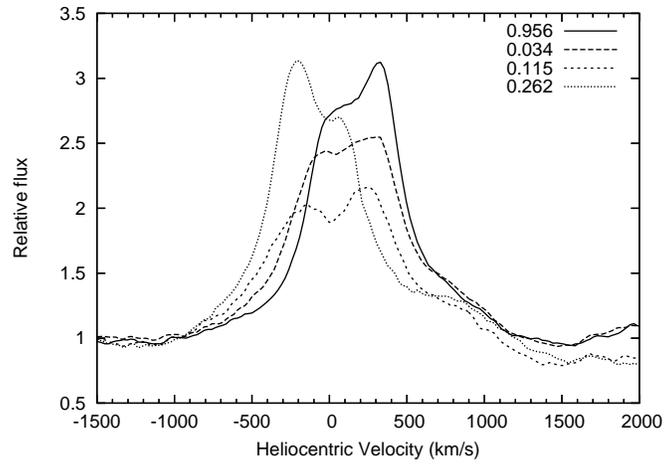}
\end{center}
\caption{Line profiles of \ion{He}{2} $\lambda 4686$. Numbers shown
in the upper right corner indicate the orbital phases.
\label{fig:HeII_lp}}
\end{figure}

To constrain the radial velocity of the compact object, we analyze the
\ion{He}{2} $\lambda 4686$ emission line, the brightest line known to
originate from the compact object. Figure~\ref{fig:HeII_lp} shows the
line corresponding profiles. The flux level is
normalized to that of the first night ($\phi$=0.956), calibrated 
using the B-band magnitudes (Figure~\ref{fig:photom}). Although the line 
profile is complex, it is obvious that the line center moves from 
the red to the blue with increasing orbital phase.

To determine the radial velocity of the \ion{He}{2} line, we calculate the
center of gravity above a certain threshold in order to discard the
broad wings. The line wings are 
stronger in the red than in the blue. We estimate
the error of the so-derived radial velocities by changing the flux
thresholds (upper and lower) used
in the calculation of the center of gravity. The results are
summarized in Table~\ref{table:data2}.

\begin{figure}
\begin{center}
\includegraphics[angle=270,scale=0.5]{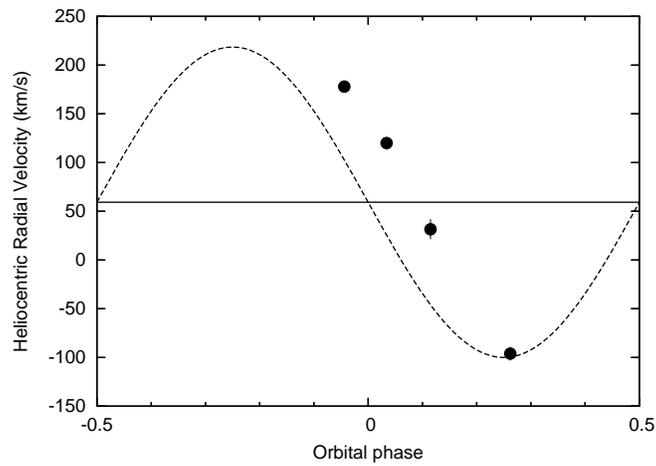}
\end{center}
\caption{Radial velocity curve of the compact object measured from
the \ion{He}{2} line. The horizontal line corresponds to the
systemic velocity of $\gamma_{\rm O}$=59.2 km s$^{-1}$ as determined
from the CCF analysis for the donor star. The dotted, sinusoidal curve
represents the best-fit obtained using the data of the fourth night 
($K_{\rm X}$=159$\pm 7$ km s$^{-1}$).}
\label{fig:sub_acc}
\end{figure}

In Figure~\ref{fig:sub_acc}, we show the radial velocities of the
compact object as measured from the \ion{He}{2} line. 
A large velocity of $\approx 150$ km s$^{-1}$ is required to fit them
with a Keplerian velocity curve. This is unlikely and probably due to
the fact that the \ion{He}{2} line was significantly affected by the
eclipse during the first three nights, as observed in previous studies
\citep{fabrika1990}.
Indeed,
the effects of the eclipse are clearly seen in Figure~\ref{fig:HeII_lp},
where the \ion{He}{2} line profile changed notably across the eclipse
of the line emitting region by the donor star. 
If we use the data of the fourth night only, together with 
a fixed systemic velocity of $\gamma_{\rm O}=59.2$ km s$^{-1}$, 
the same value of $\gamma_{\rm x}$ for the donor
star, we obtain K$_{\rm X}$=159$\pm7$ km s$^{-1}$. This value is
consistent with previous results.

\citet{fabrika1990} reported $K_{\rm X}=175\pm20$ km s$^{-1}$, using
the \ion{He}{2} $\lambda 4686$ line observed in the precession phase of
$0.9 \le \psi \le 0.1$, but outside of the eclipse. 
\citet{fabrika1997} constructed precessional and orbital
radial velocity curves of the \ion{He}{2} line using additional
spectral data.  They found $K_{\rm X}=176\pm15$ km s$^{-1}$ for the
same precessional phase of $0.9 \le \psi \le0.1$, while
\citet{gies2002b} 
used \ion{C}{2} $\lambda7231, 7236$
blended lines and derived $K_{\rm X}=162\pm29$ km s$^{-1}$. For 
consistency, we adopt $K_{\rm X}$=168$\pm$10 km s$^{-1}$ in this
paper, the average between these three studies and our own
estimate of the $K_{\rm X}$.  Note that \citet{hillwig2004} and
\citet{hillwig2008} adopted the same value, 168$\pm$18 km s$^{-1}$, as
the average between two studies, \citet{gies2002b} and
\citet{fabrika1990}.

\section{DISCUSSION}

\subsection{Review of the Dynamical Mass Determination of SS~433}

In this subsection, we review recent work on the dynamical
determination of the mass function of SS~433 from measurements of
$K_{\rm X}$ and $K_{\rm O}$, which we compare with our results.

\begin{enumerate}

\item \citet{gies2002b} interpreted that an absorption feature in the
strong emission line of \ion{He}{1} $\lambda 6678$ may originate from
the donor star, using the spectra taken at the KPNO 0.9 m
telescope. They derived $K_{\rm O}$ = 126$\pm$26 km
s$^{-1}$. Combining this values with $K_{\rm X}$=175$\pm$20 km s$^{-1}$
\citep{fabrika1990}, they determined $M_{\rm x}$=16$\pm$6 \solarmass\
and $M_{\rm O}$=23$\pm$8 \solarmass.

\item \citet{gies2002a} measured the radial velocity of the donor star
using the CCF technique for the first time, based on data taken with
the 2.7 m telescope of the University of Texas McDonald. For 
the 4060--4750 \AA\ range,
they calculated CCFs between different SS~433 spectra. They obtained
$K_{\rm O}$=100$\pm$15~km~s$^{-1}$ and $\gamma_{\rm O}$=$-$44$\pm$9~km~
s$^{-1}$. Together with $K_{\rm x}$=175$\pm$20 km s$^{-1}$
\citep{fabrika1990}, they concluded $M_{\rm X}$=11$\pm$5 \solarmass\
and $M_{\rm O}$=19$\pm$7 \solarmass.

\item \citet{hillwig2004} applied a CCF analysis of Region~1
(as defined in our paper) to spectra taken at the KPNO 4 m telescope (in
this epoch, Region~1 did not contain strong emission lines,
unlike in the case of \citealt{hillwig2008}). They used the same reference star
HD~9233 (type A4 Iab) as we used in this paper, and derived $K_{\rm
O}$=45$\pm$6 km s$^{-1}$ and $\gamma_{\rm O}$=65$\pm$3 km
s$^{-1}$. Together with $K_{\rm x}$=168$\pm$18 km s$^{-1}$ (the
average value of \citealt{fabrika1990} and \citealt{gies2002b}), they
determined $M_{\rm X}$=2.9$\pm$0.7 \solarmass\ and $M_{\rm
O}$=10.9$\pm$3.1\solarmass.

\item \citet{cherepashchuk2005} examined Doppler shifts of 22
absorption lines in the 4200--5300 \AA\ range relative to the
laboratory frame, using spectra taken at the SAO 6 m
telescope. They obtained $K_{\rm O}$=132$\pm$9 km s$^{-1}$ and
$\gamma_{\rm O}$=14 km s$^{-1}$. Combining these values with $K_{\rm x}$=175 km
s$^{-1}$ \citep{fabrika1990}, they determined $M_{\rm X} \approx$ 18
\solarmass\ and $M_{\rm O} \approx$24 \solarmass. They noted that this
radial velocity semi-amplitude is probably an upper limit, because the
      heating of the donor star increases the observed amplitude.

\item \citet{barnes2006} observed SS~433 with the Calar Alto
Observatory 3.5 m telescope, and the Observatory del Roque de Los
Muchachos 2.5 m and 4.2 m telescope. They cross-correlated the SS~433
spectra in the 4500--4630 \AA\ range with those of HD~9233, using the
radial velocity given in \citet{hillwig2004}. They obtained
$K_{\rm O}$=69$\pm$4 km s$^{-1}$ and $\gamma_{\rm O}$=$-$53$\pm$3 km
s$^{-1}$.
But, their observations were performed when the accretion disk was
close to an edge-on orientation, where the
outflowing material produces strong absorption lines. Note that
they assumed a velocity for HD~9233 of $-$34 km s$^{-1}$
\citep{hillwig2004}. Since this star is velocity variable, its
velocity might have been different at the time of the observations, and
the systemic velocity $\gamma_{\rm 0}$ has an additional factor of uncertainty.

\item \citet{hillwig2008} made a CCF analysis in Regions~2 and 3 using
the Gemini data. They obtained $K_{\rm O}$=58.2$\pm$3.1 km s$^{-1}$
and $\gamma_{\rm O}$=73$\pm$2 km s$^{-1}$, which together with $K_{\rm
X}$=168$\pm$18 km s$^{-1}$ (the average of \citealt{fabrika1990} and
\citealt{gies2002b}) leads to $M_{\rm X}$=4.3$\pm$0.8
\solarmass\ and $M_{\rm O}$=12.3$\pm$3.3 \solarmass.

\end{enumerate}

\subsection{Constraints on the Mass of the Compact Object}

Once the amplitudes of the radial velocity curves of both the donor star
and the compact object are known, one can deduce the mass of each
component.  By fitting the radial velocity curve of the donor star
obtained from the CCF analysis with a Keplerian curve (i.e., without
consideration for the heating effects), we derived $K_{\rm
O}=58.3\pm3.8$ km s$^{-1}$, which is consistent with the value
obtained by \citet{hillwig2008}. This is not surprising, since we
use the same Gemini data for our analysis in addition to the Subaru
data, although we restrict the wavelength range to Region~2 only, in
order to
avoid systematic effects from emission components in the
absorption line profiles. We adopt the amplitude of the
radial velocity of the compact object of $K_{\rm X}$=168$\pm$10 km s$^{-1}$
and conclude the mass of the donor star and
compact object to be $M_{\rm O}=12.4 \pm 1.9$ \solarmass\ and
$M_{\rm X}=4.3 \pm 0.6$ \solarmass, respectively. The
corresponding mass ratio is $q=0.35$. Again, these values should be
taken as upper limits only if we consider the heating effect, as
discussed in Section~\ref{sec:heating}.

By taking into account the heating effects, we
derive lower limits on the masses, since we assumed a real  
radial velocity amplitude of the donor star of
$K_{\rm O} = 40 \pm 5$ km s$^{-1}$. This leads to lower 
masses of $M_{\rm O} = 
10.4^{+2.3}_{-1.9}$ \solarmass\ and $M_{\rm X} = 2.5^{+0.7}_{-0.6}$ 
\solarmass.
We thus conclude that the compact object in SS~433 is most
likely a low mass black hole. However, the possibility of a massive
neutron star cannot be firmly ruled out at present, given the fact
that a neutron stars could have masses of up to 3
\solarmass\, as inferred from both theory \citep{lattimer2007} and
observations \citep{freire2008a}.

\section{CONCLUSIONS}

\begin{enumerate}

\item To study the radial velocity curve of the mass donor star in
SS~433, we obtained high quality optical spectra with Subaru/FOCAS,
covering the orbital phase of $\phi=0.96-0.26$. We combine these
      observations with the
Gemini data reported by \citet{hillwig2008} to analyze the largest
set of the best quality spectra observed right now from this source. This
allows us to study in detail the behavior of the ``weak'' absorption
lines from the donor surface, which are least affected by the emission
components from the surroundings of the donor star.

\item We demonstrate that the selection of the spectral region is critical
for the cross correlation function (CCF) analysis. We adopt the
4740--4840 \AA\ range (Region~2) for this study, where only ``weak''
absorption lines from the surface of the donor star are present. If we
instead use the 4490--4630 \AA\ range (Region~1), which contains many
``strong'' absorption lines associated with emission components, we
obtain a significantly smaller velocity amplitude than Region~2.

\item From the Subaru and Gemini CCF results (Region~2), we determine
the amplitude of the radial velocity curve of the donor star to be
58.3$\pm$3.8 km s$^{-1}$. Together with the radial velocity of the
compact object, 168$\pm$10 km s$^{-1}$, we derive masses of the
donor star and the compact object of $M_{\rm O}=12.4 \pm 1.9$
\solarmass\ and $M_{\rm X}=4.3 \pm 0.6$ \solarmass,
respectively. These values should be taken as {\it upper limits},
because of the heating of the donor star by the compact object.

\item We calculated average absorption line profiles 
for the strong and weak lines separately, each line using 8 individual
      lines. The position of the line centers of the average absorption
      line agree within our CCF results. 
      Prominent emission
components are observed in the strong lines, indicating that the
heating effects are important for a proper interpretation.

\item We construct a simple model where we take into account 
the UV heating effects on the donor star surface and on its envelope,
      and where  also consider 
the emission from the gas stream. The model reproduces well the
emission components and absorption lines in the average line profiles,
both in intensity and radial velocity variations. These results
indicate that the heating could have a significant impact on the
estimate of the real radial velocity of the donor star,
which may be as low as $K_{\rm O} = 40 \pm 5$~km~s$^{-1}$.
We then estimate the masses of the components 
as $M_{\rm O} = 10.4^{+2.3}_{-1.9}$~\solarmass\ and $M_{\rm X} 
= 2.5^{+0.7}_{-0.6}$~\solarmass.

\item The final constraints for the compact object mass are 1.9
\solarmass\ $\leq M_{\rm X} \leq$ 4.9 \solarmass, where the lower and
upper limits are inferred from the modeling of the average absorption line
profiles and from the CCF analysis, respectively. We conclude that the
compact object in SS~433 is most likely a low mass black hole, although
the possibility of a massive neutron star cannot be firmly excluded.

\end{enumerate}

\acknowledgments

We thank the observatory staff of the Subaru telescope, in particular
the support astronomer Dr.\ Takashi Hattori, for their help on our
observation run, the scheduling, and for useful advice on the reduction of the
FOCAS data. We also thank Dr.\ E.\ Chentsov for the very useful
comments, and the referee, Prof.\ Douglas Gies, for his careful
reading of the manuscript that helped to improve the clarity of the
paper. 
We are grateful to Dr. Dominikus Heinzeller, who gave extensive
suggestions to the manuscript.
KK and YU greatly appreciate the warm hospitality of staffs in
the SAO during their visits in 2007 and 2008. This work was partly
supported by the Grant-in-Aid for JSPS Fellows for young researchers
(KK), by the Grant-in-Aid for the Global COE Program ``The Next
Generation of Physics, Spun from Universality and Emergence'' from
from the Ministry of Education, Culture, Sports, Science and
Technology (MEXT) of Japan, and by Russian RFBR grants 07-02-00909,
07-02-00630, and 09-02-00163.

\bibliographystyle{apj}
\bibliography{biblist}

\clearpage

\clearpage

\begin{deluxetable}{cccc}
\tablecaption{Log of Observations}
\label{table:obslog}
\tablehead{
\colhead{Date}
&\colhead{HJD$-$2,450,000}
&\colhead{Total exposure}
&\colhead{Frames}\\
\colhead{}
&\colhead{(Mid time)}
&\colhead{(sec)}
&\colhead{}
}
\startdata
\multicolumn{4}{l}{\bf{Subaru}}\nl
2007 Oct 06&4379.76&6000&5\nl
2007 Oct 07&4380.78&8400&7\nl
2007 Oct 08&4381.84&5484&5\nl
2007 Oct 10&4383.77&8400&7\nl
\hline
\multicolumn{4}{l}{\bf{Gemini}}\nl
2006 Jun 07&3893.99&5400&3\nl
2006 Jun 08&3895.04&9000&5\nl
2006 Jun 09&3895.96&16200&9\nl
2006 Jun 10&3897.02&12600&7\nl
2006 Jun 11&3898.01&9000&5\nl
2006 Jun 12&3899.03&7200&4\nl
2006 Jun 13&3900.03&7200&4\nl
\enddata
\end{deluxetable}

\begin{deluxetable}{ccccc}
\tablecaption{Radial Velocity of the Donor Star Determined from the
 CCF (Region~2) Together with Precessional and Orbital Phases
\label{table:data}}
\tablehead{\colhead{Date}&\colhead{Observatory}&\colhead{$\psi$}&\colhead{$\phi$}&\colhead{$V_{r}$}\\
\colhead{(HJD$-$2,450,000)}&\colhead{}&\colhead{}&\colhead{}&\colhead{(${\rm km\,s^{-1}}$)}}
\startdata
4379.76&Subaru&0.018&0.956&52$\pm$7\nl
4380.78&Subaru&0.024&0.034&65$\pm$7\nl
4381.84&Subaru&0.031&0.115&84$\pm$7\nl
4383.77&Subaru&0.043&0.262&119$\pm$7\nl
3893.99&Gemini&0.022&0.838&9$\pm$8\nl
3895.04&Gemini&0.029&0.918&39$\pm$8\nl
3895.96&Gemini&0.034&0.988&47$\pm$8\nl
3897.02&Gemini&0.041&0.069&73$\pm$8\nl
3898.01&Gemini&0.047&0.145&106$\pm$9\nl
3899.03&Gemini&0.053&0.223&132$\pm$9\nl
3900.03&Gemini&0.060&0.300&128$\pm$8\nl
\enddata
\tablecomments{The errors are 1$\sigma$ standard errors from the CCF analysis.}
\end{deluxetable}

\begin{deluxetable}{ccccc}
\tablecaption{Radial Velocity of the \ion{He}{2} line\label{table:data2}}
\startdata
\tablehead{\colhead{Date}&\colhead{Observatory}&\colhead{$\psi$}&\colhead{$\phi$}&\colhead{$V_{r}$}\\
\colhead{(HJD$-$2,450,000)}&\colhead{}&\colhead{}&\colhead{}&\colhead{(${\rm km\,s^{-1}}$)}}
4379.76&Subaru&0.018&0.956&178$\pm$2\\
4380.78&Subaru&0.024&0.034&120$\pm$1\\
4381.84&Subaru&0.031&0.115&32$\pm$10\\
4383.77&Subaru&0.043&0.262&$-$96$\pm$7\\
\enddata
\tablecomments{The errors are estimated systematic errors (see text).}
\end{deluxetable}
\end{document}